\documentclass[12pt]{iopart}

\usepackage{iopams}
\usepackage{verbatim} %mehrzeilige Kommentare
\usepackage[utf8]{inputenc}

\usepackage{graphicx}
\usepackage[colorlinks=true,citecolor=blue]{hyperref}
\usepackage{units}

\newcommand{\ket}[1]{|#1\rangle}
\renewcommand{\vec}[1]{\mathbf{#1}}
\newcommand{\text}[1]{\mathsf{#1}}

\def\up{\uparrow}
\def\down{\downarrow}

\newcommand{\kBT}{k_\mathsf{B}T}

\begin{document}

\title{Fractional Josephson effect in a quadruple quantum dot}

\author{Bj\"orn Sothmann, Jian Li and Markus B\"uttiker}
\address{D\'epartement de Physique Th\'eorique, Universit\'e de Gen\`eve, CH-1211 Gen\`eve 4, Switzerland}
\eads{\mailto{bjorn.sothmann@unige.ch},\mailto{jian.li@unige.ch},\mailto{markus.buttiker@unige.ch}}

\date{\today}

\begin{abstract}
A double quantum dot coupled to an $s$-wave superconductor and subject to an inhomogeneous magnetic field can host a pair of zero-energy Majorana fermions when the dot properties are tuned appropriately.
Here, we demonstrate the possibility to generate a fractional $4\pi$ Josephson effect in two such double dots tunnel-coupled to each other. We discuss the robustness of this effect with respect to perturbations away from the special point in parameter space where the uncoupled double dots host Majorana fermions. We demonstrate the possibility to generate Josephson effects with a period of $8\pi$ and $12\pi$ in strongly-coupled double dots.
\end{abstract}

% 74.45.+c 	Proximity effects; Andreev reflection; SN and SNS junctions
% 74.50.+r 	Tunneling phenomena; Josephson effects
% 74.20.Mn 	Nonconventional mechanisms 
% 74.20.Rp 	Pairing symmetries (other than s-wave) 
% 73.63.-b 	Electronic transport in nanoscale materials and structures
% 73.63.Kv 	Quantum dots 
% 73.23.Hk 	Coulomb blockade; single-electron tunneling 
% 74.78.-w 	Superconducting films and low-dimensional structures
% 74.78.Na 	Mesoscopic and nanoscale systems 

\submitto{\NJP}

\maketitle

\section{\label{sec:introduction}Introduction}
Currently, there is a rapidly increasing interest in the possibility to generate Majorana fermions in condensed-matter systems~\cite{alicea_new_2012,beenakker_search_2013}. 
This is triggered by the fact that Majorana fermions exhibit non-Abelian braiding properties and offer the possibility to realize topological qubits that are immune to local perturbations for quantum computing~\cite{alicea_non-abelian_2011,flensberg_non-abelian_2011,van_heck_coulomb-assisted_2012}.

By now, various theoretical proposals of how to realize Majorana fermions exist, e.g., by quasiparticles in the $\nu=5/2$ fractional quantum Hall state~\cite{moore_nonabelions_1991}, as edge states of a one-dimensional wire made from a $p$-wave superconductor~\cite{kitaev_unpaired_2001} or in vortices of a $p$-wave superconductor~\cite{das_sarma_proposal_2006}. Unfortunately, all of these proposal are very hard to realize experimentally. Recently, several more feasible setups have been suggested where the surface states of a three-dimensional topological insulator or edge states of a quantum spin Hall insulator are coupled to an $s$-wave superconductor~\cite{fu_superconducting_2008,fu_josephson_2009}. In addition, it was shown that a one-dimensional $p$-wave superconductor can effectively be induced in a single-channel nanowire with strong spin-orbit interaction subject to an external magnetic field and in proximity with a standard $s$-wave superconductor~\cite{lutchyn_majorana_2010,oreg_helical_2010}.

Majorana fermions give rise to a number of characteristic transport signatures. First of all, they lead to a zero-bias peak of $2e^2/h$ in the differential conductance of a weakly coupled normal metal probe~\cite{law_majorana_2009,flensberg_tunneling_2010}. In addition, Majorana fermions also give rise to a fractional Josephson effect where the supercurrent is $4\pi$ periodic in the phase difference between the superconductors instead of being $2\pi$ periodic~\cite{kitaev_unpaired_2001}. The fractional Josephson effect is a consequence of the crossing of Andreev bound state energies as a function of the phase difference. It has by now been studied theoretically in Josephson junctions where Majorana fermions are realized using quantum spin Hall insulator edge states~\cite{fu_josephson_2009,badiane_nonequilibrium_2011,beenakker_fermion-parity_2013}, surface states of a three-dimensional topological insulator~\cite{ioselevich_anomalous_2011} and nanowires~\cite{jiang_unconventional_2011}. The effects of many-channel wires~\cite{law_robustness_2011}, finite wire length~\cite{san-jose_ac_2012} and Coulomb charging have been investigated~\cite{van_heck_coulomb_2011}.
The possibility to observe a fractional Josephson effect due to Landau-Zener tunneling in realistic systems where imperfections in the system turn the level crossing into an avoided crossing has been discussed~\cite{dominguez_dynamical_2012,pikulin_phenomenology_2012}. Landau-Zener tunneling has also been shown to yield a fractional Josephson effect even in Josephson junctions without Majorana fermions~\cite{sau_possibility_2012}.

Nanowire setups that are supposed to host Majorana fermions at their ends have by now been realized by several groups~\cite{mourik_signatures_2012,das_zero-bias_2012,deng_anomalous_2012,finck_anomalous_2013,churchill_superconductor-nanowire_2013}. These experiments have indeed shown zero-bias conductance peaks in agreement with theoretical predictions. However, there exist other mechanisms like interference effects in the presence of disorder~\cite{bagrets_class_2012} and weak antilocalization~\cite{pikulin_zero-voltage_2012} that can yield similar zero-bias anomalies.
Additionally, recent experiments have shown the occurrence of fractional Shapiro steps in Josephson junctions from InSb nanowires~\cite{rokhinson_fractional_2012} as well as unconventional behaviour of a superconductor-topological insulator Josephson junction~\cite{williams_unconventional_2012}.

One of the biggest issues in clearly identifying signatures of Majorana fermions is the fact that in experiments the nanowire is buried beneath a superconducting electrode. Hence, its properties are not accessible to direct characterization. This problem can be circumvented by alternative setups based on quantum dots~\cite{sau_realizing_2012,leijnse_parity_2012,fulga_adaptive_2013,wright_localized_2013,brunetti_anomalous_2013}. The simplest realization of such a system consists of a double quantum dot coupled to an $s$-wave superconductor and subject to an inhomogenous magnetic field~\cite{leijnse_parity_2012}. The presence of Majorana fermions in such a system is indicated by characteristic features in the zero-bias conductance in analogy to the Majorana setups realized in nanowires.

Here, we demonstrate the possibility to generate a fractional Josephson effect in a setup of two double dots and analyse the robustness of the effect with respect to level detunings and Coulomb interactions. Compared to a conductance measurement, the Josephson current offers the advantage of being a non-invasive probe as it does not require to couple any additional normal electrode to the system.

Using quantum dots to investigate Majorana fermions offers a number of advantages. First of all, quantum dots are well-controlled systems with parameters like level positions that can be easily tuned in experiment. In addition, very similar setups have already been realized to act as Cooper pair splitter~\cite{hofstetter_cooper_2009,herrmann_carbon_2010,hofstetter_finite-bias_2011,das_high-efficiency_2012,schindele_near-unity_2012}. Furthermore, supercurrents through quantum dots have been observed in a number of experiments as well~\cite{van_dam_supercurrent_2006,cleuziou_carbon_2006,grove-rasmussen_kondo_2007,winkelmann_superconductivity_2009,eichler_tuning_2009,kanai_electrical_2010,lee_zero-bias_2012,kim_transport_2013}. From a theoretical perspective, a double quantum dot represents the shortest Kitaev chain consisting of two links only. Hence, it is the simplest possible Majorana setup. Finally, quantum dots are an ideal playground to investigate the interplay between Majorana physics and strong Coulomb interactions.

The paper is organized as follows. We introduce our model in Sec.~\ref{sec:model} and discuss how to evaluate the Josephson current from the quantum dot spectrum in Sec.~\ref{sec:Josephson}. In Sec.~\ref{sec:energies}, we discuss the eigenenergies which are directly related to the Josephson current. Finally, we discuss the limit of strong coupling between the double dots where additional crossings occur in Sec.~\ref{sec:strong}. We finish with conclusions given in Sec.~\ref{sec:conclusions}.

\section{\label{sec:model}Model}
\begin{figure}
	\centering\includegraphics[width=\columnwidth]{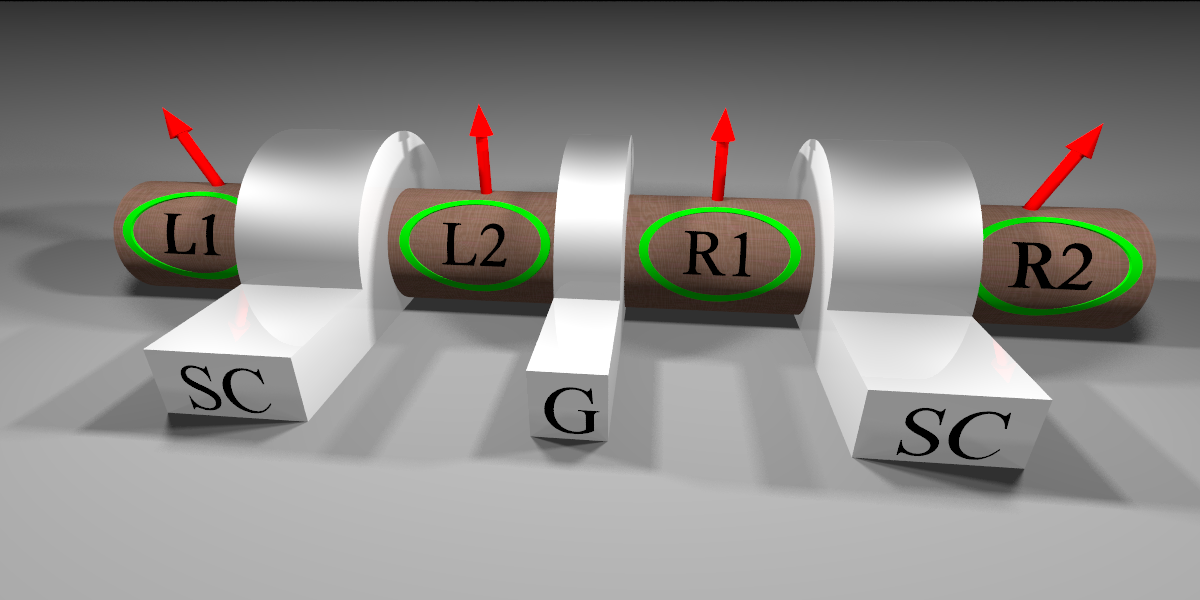}
	\caption{\label{fig:model}Sketch of the quadruple quantum dot. Quantum dots L1 and L2 as well as quantum dots R1 and R2 form double quantum dots coupled to superconducting electrodes with phases $\Phi_\text{L}$ and $\Phi_\text{R}$, respectively. The two double quantum dots are coupled to each other by a tunnel barrier controlled by the gate voltage applied to $G$. Red arrows indicate the inhomogenous magnetic field.}
\end{figure}

We consider a system of four single-level quantum dots as shown in \fref{fig:model}. The quantum dots form two double dots, $r=\mathsf{L,R}$, each coupled to a grounded superconductor. As we are interested in subgap transport only, we focus on the limit of infinite superconducting gaps.
The whole system is subject to a magnetic field that lifts spin degeneracy. In order to have both, direct tunneling between the dots $i=1,2$ within a double dot and a superconducting proximity effect caused by crossed Andreev reflections, we assume the magnetic field acting on the two dots to enclose a finite angle $\alpha_r$. By changing this angle, the relative strength between direct tunneling and the proximity effect can be tuned~\cite{leijnse_parity_2012}.
Finally, the two double dots are connected to each other via a tunnel coupling between dot $i=2$ in the left and dot $i=1$ in the right double dot.

\subsection{\label{ssec:HEff}Effective Hamiltonian}
The system is characterized by the total Hamiltonian
\begin{equation}\label{eq:H}
 	H=\sum_r H_{r,\mathsf{eff}}+H_\mathsf{tun}+H_\mathsf{int}.
\end{equation}
The effective dot Hamiltonian that describes the double quantum dot $r$ is given by
\begin{equation}\label{eq:Heff}
	H_{r,\mathsf{eff}}=\sum_i \varepsilon_{ri} n_{ri}+t_r c_{r1}^\dagger c_{r2}+\Delta_r\rme^{\rmi\Phi_r} c_{r1}^\dagger c_{r2}^\dagger+\mathsf{H.c.},
\end{equation}
cf.~\ref{app:Heff} for a detailed derivation. Here, $\varepsilon_{ri}$ denotes the energy of the single level relevant for transport in the two dots. Direct tunneling between the two dots is characterized by $t_r$. The superconducting proximity effect is described by the induced gap $\Delta_r$ which in our system is proportional to the tunnel coupling between the double dot and the superconductor. Finally, $\Phi_r$ denotes the phase of the superconducting order parameter in the electrode coupled to double dot $r$.

The coupling between the two double dots is described by the tunneling Hamiltonian
\begin{equation}
	H_\mathsf{tun}=t c_{\mathsf{L}2}^\dagger c_{\mathsf{R}1}+\mathsf{H.c.},
\end{equation}
where the tunnel matrix element $t$ can be controlled by a gate voltage.
Finally, interactions are captured by the interaction Hamiltonian,
\begin{equation}
	H_\mathsf{int}=\sum_{rir'i'} U_{rir'i'} n_{ri} n_{r'i'},
\end{equation}
where the sum runs over all pairs of different quantum dots and double counting is excluded.

\subsection{\label{ssec:Majorana}Majorana representation}
\begin{figure}
	\begin{center}
	\includegraphics[width=.45\columnwidth]{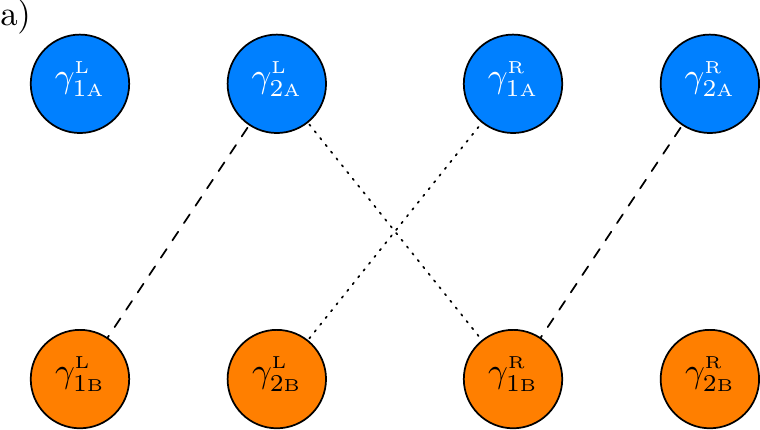}
	\hfill
	\includegraphics[width=.45\columnwidth]{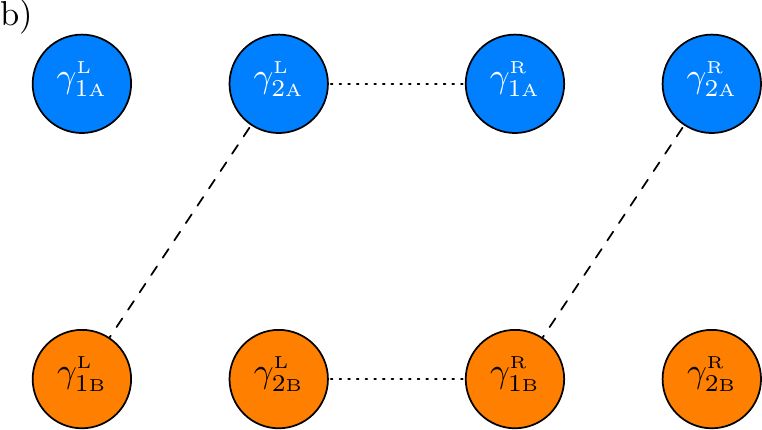}
	\newline
	\newline
	\includegraphics[width=.45\columnwidth]{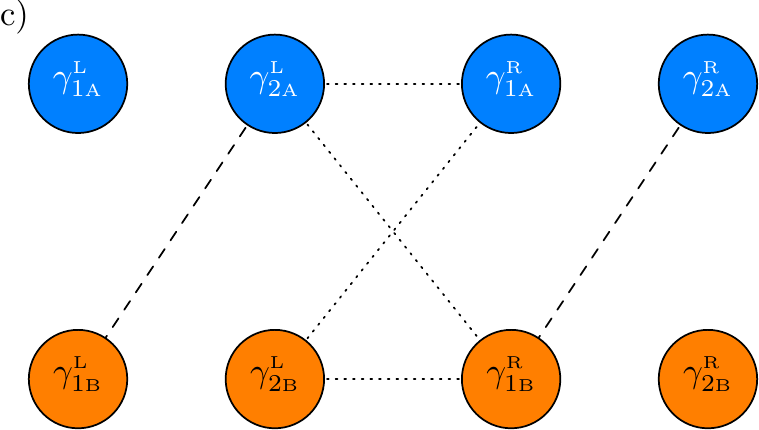}
	\end{center}
	\caption{\label{fig:Majorana}Representation of the quadruple dot in terms of Majorana fermions. Each column denots a quantum dot that can be described in terms of two Majorana fermions $\gamma^r_{iA}$ and $\gamma^r_{iB}$. Dashed lines indicate couplings between Majoranas arising from tunneling and proximity effect within a double dot. The dotted lines indicate couplings due to the tunneling between the double dots. In a) the phase difference is $\Delta\Phi=2n\pi$, in b) it is $\Delta\Phi=(2n+1)\pi$ while c) shows a generic situation.}
\end{figure}

We now express the Hamiltonian~\eref{eq:H} in terms of Majorana fermions. To this end, we introduce the Majorana operators
\begin{eqnarray}\label{eq:Majorana}
	\gamma_{i\mathsf{A}}^r&=\rme^{-\rmi\Phi_r/2}c_{ri}+\rme^{\rmi\Phi_r/2}c_{ri}^\dagger,\\
	\gamma_{i\mathsf{B}}^r&=-\rmi\rme^{-\rmi\Phi_r/2}c_{ri}+\rmi\rme^{\rmi\Phi_r/2}c_{ri}^\dagger,
\end{eqnarray}
which satify $(\gamma_{i\mathsf{A/B}}^{r})^\dagger=\gamma_{i\mathsf{A/B}}^r$. The inverse transformation reads
\begin{eqnarray}
	c_{ri}&=\frac{1}{2}\rme^{\rmi\Phi_r/2}\left(\gamma^r_{i\mathsf{A}}+\rmi\gamma^r_{i\mathsf{B}}\right),\\
	c_{ri}^\dagger&=\frac{1}{2}\rme^{-\rmi\Phi_r/2}\left(\gamma^r_{i\mathsf{A}}-\rmi\gamma^r_{i\mathsf{B}}\right).
\end{eqnarray}
In terms of the Majorana operators, the effective dot Hamiltonian~\eref{eq:Heff} becomes
\begin{equation}\label{eq:HMajorana}
	H_{r,\mathsf{eff}}=\frac{\rmi}{2}\left[-(t_r+\Delta_r)\gamma_{1\mathsf{B}}^r\gamma_{2\mathsf{A}}^r+(t_r-\Delta_r)\gamma_{1\mathsf{A}}^r\gamma_{2\mathsf{B}}^r+\sum_i\varepsilon_{ri}\gamma_{i\mathsf{A}}^r\gamma_{i\mathsf{B}}^r\right],
\end{equation}
where we dropped an irrelevant constant. From~\eref{eq:HMajorana}, we directly read off that the double dot hosts a pair of zero-energy excitations at the special point $t_r=\Delta_r$, $\varepsilon_i=0$ described by the Majorana operators $\gamma_{1\mathsf{A}}^r$ and $\gamma_{2\mathsf{B}}^r$.

Expressed in terms of Majorana fermions, the tunnel coupling and the Coulomb interaction become
\begin{equation}
	H_\mathsf{tun}=\frac{\rmi t}{2}\left[\sin\frac{\Delta\Phi}{2}\left(\gamma^\mathsf{L}_{2\mathsf{A}}\gamma^\mathsf{R}_{1\mathsf{A}}+\gamma^\mathsf{L}_{2\mathsf{B}}\gamma^\mathsf{R}_{1\mathsf{B}}\right)+\cos\frac{\Delta\Phi}{2}\left(\gamma^\mathsf{L}_{2\mathsf{A}}\gamma^\mathsf{R}_{1\mathsf{B}}-\gamma^\mathsf{L}_{2\mathsf{B}}\gamma^\mathsf{R}_{1\mathsf{A}}\right)\right],
\end{equation}
and
\begin{equation}
	H_\mathsf{int}=\sum_{rir'i'}\frac{U_{rir'i'}}{4}\left[1+\rmi\left(\gamma^r_{i\mathsf{A}}\gamma^r_{i\mathsf{B}}+\gamma^{r'}_{i'\mathsf{A}}\gamma^{r'}_{i'\mathsf{B}}\right)-\gamma^r_{i\mathsf{A}}\gamma^r_{i\mathsf{B}}\gamma^{r'}_{i'\mathsf{A}}\gamma^{r'}_{i'\mathsf{B}}\right],
\end{equation}
respectively. Here $\Delta\Phi=\Phi_\mathsf{R}-\Phi_\mathsf{L}$ denotes the phase difference between the two superconductors.

A schematic representation of the quadruple dot in terms of its Majorana degrees of freedom for different values of the phase difference $\Delta\Phi$ is shown in \fref{fig:Majorana}.

\section{\label{sec:Josephson}Josephson current}
The Josephson current through the quantum dots is given by the derivative of the free energy with respect to the phase difference between the superconductors~\cite{beenakker_three_1992},
\begin{equation}\label{eq:JJos1}
	J_\mathsf{jos}=\frac{2e}{\hbar}\frac{\partial F}{\partial \Delta\Phi}.
\end{equation}
As the free energy is given by $F=-\kBT\log \sum_i \rme^{-E_i/\kBT}$ where $E_i$ denote the many-body eigenenergies of the system, at low temperatures, $\kBT\ll E_i-E_j$, the Josephson current is essentially determined by the ground state energy $E_0$,
\begin{equation}
	J_\mathsf{jos}=\frac{2e}{\hbar}\frac{\partial E_0}{\partial \Delta\Phi}.
\end{equation}
We remark that $E_0$ is a many-body ground state energy, i.e., it is the lowest eigenenergy of the system. This is in contrast to a description in the Bogoliubov-de Gennes picture where the excitation energies close to zero energy determine the Josephson current. Since the Josephson current is completely determined by the spectrum of the quantum dot, in the following, we will discuss the properties of the eigenenergies rather than the Josephson current itself.

\section{\label{sec:energies}Eigenenergies}
We are now going to discuss the eigenenergies of the quantum dot. We focus on the situation where $t_r=\Delta_r$. As soon as this condition is relaxed, the system will no longer exhibit a fractional Josephson effect. This is in contrast to a nanowire setup where a fractional Josephson effect can occur even when $t_r\neq\Delta_r$ since the induced splitting of the zero-energy states is exponentially small in the length of the wire.
First of all, we discuss the ``sweet spot'' in parameter space where in addition to $t_r=\Delta_r$ we have $\varepsilon_{ri}=0$ and $U_{rir'i'}=0$. We then turn to the effect of detuned levels and analyse the influence of Coulomb interactions.

\subsection{\label{ssec:sweet}Sweet spot}
\begin{figure}
	\centering\includegraphics[width=.6\columnwidth]{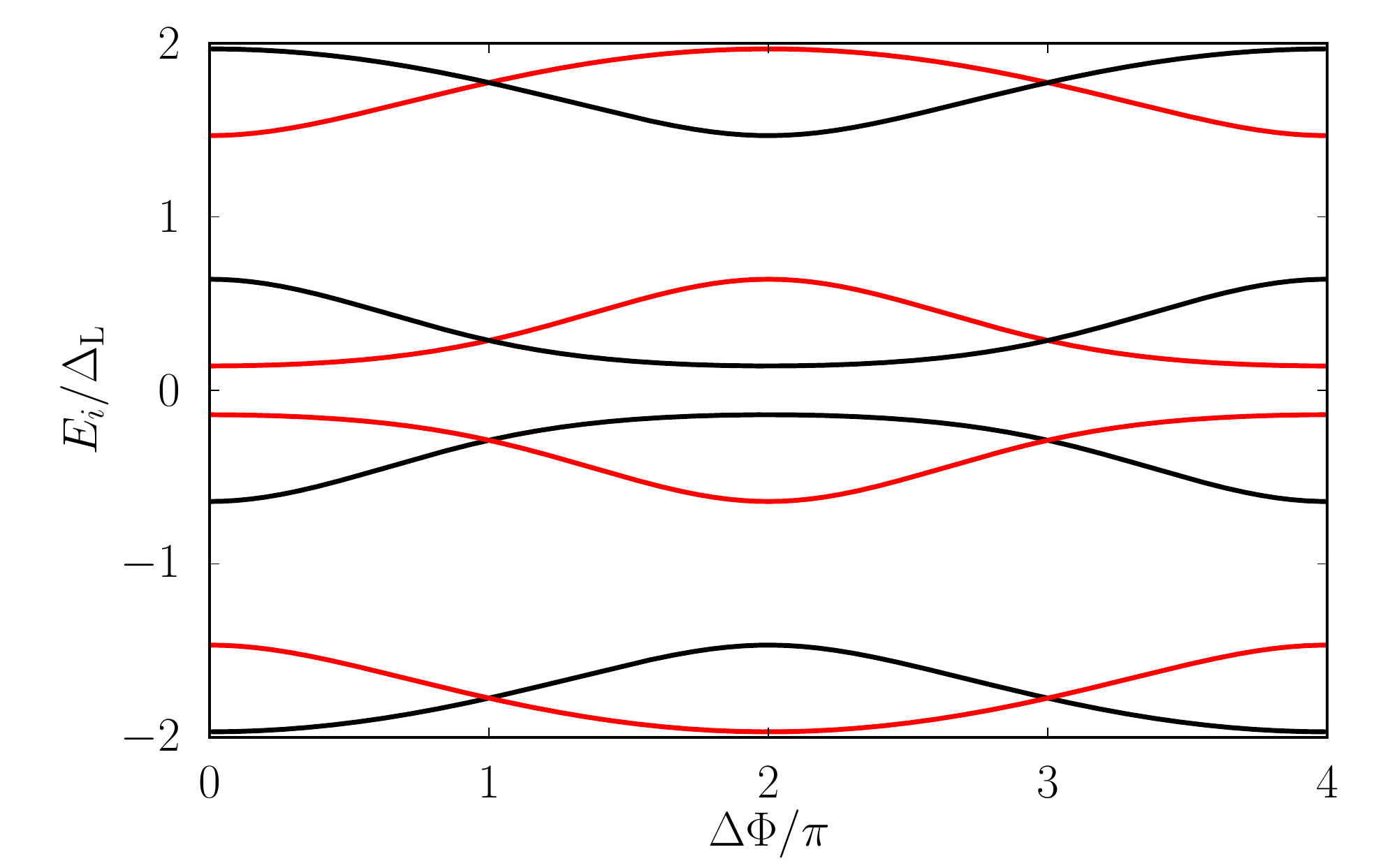}
	\caption{\label{fig:energies}Eigenenergies as a function of the phase difference $\Delta\Phi$. Parameters are $\Delta_\mathsf{R}=0.7\Delta_\mathsf{L}$, $\varepsilon_{ri}=0$, $t_r=\Delta_r$, $t=0.5\Delta_\mathsf{L}$, $U_{ri,r'i'}=0$. For each eigenenergy, there are two degenerate states with an even and odd number of electrons on the quantum dots, respectively. Note that we discuss the many-body eigenenergies, i.e., the ground state properties are determined by the state with lowest energy.}
\end{figure}

At the sweet spot, $t_r=\Delta_r$, $\varepsilon_{ri}=0$ and $U_{ri,r'i'}=0$, analytical expressions for the eigenenergies can be obtained. However, as they are lengthy and do not provide much insight it is more convenient to discuss the eigenenergies obtained via degenerate perturbation theory in the coupling $t$ between the double dots. Up to second order in $t$, we find
\begin{eqnarray}
	\label{eq:E0}
	E_{0\pm}=&-|\Delta_\text{L}+\Delta_\text{R}|
	\pm\frac{t}{2}\cos\frac{\Delta\Phi}{2}\\
	&-t^2\frac{(\Delta_\text{L}+\Delta_\text{R})^2(1-\cos\Delta\Phi)+\Delta_\text{L}\Delta_\text{R}(1+\cos\Delta\Phi)}{16\Delta_\text{L}\Delta_\text{R}(\Delta_\text{L}+\Delta_\text{R})},\nonumber\\
% 	&\pm t^3\frac{\Delta_\text{L}^2+\Delta_\text{R}^2}{32\Delta_\text{L}^2\Delta_\text{R}^2}\cos\frac{\Delta\Phi}{2}\left(\cos\Delta\Phi-1\right),\\
	E_{1\pm}=&-|\Delta_\text{L}-\Delta_\text{R}|
	\pm\frac{t}{2}\cos\frac{\Delta\Phi}{2}\\
	&-t^2\frac{(\Delta_\text{L}-\Delta_\text{R})^2(1-\cos\Delta\Phi)-\Delta_\text{L}\Delta_\text{R}(1+\cos\Delta\Phi)}{16\Delta_\text{L}\Delta_\text{R}(\Delta_\text{L}+\Delta_\text{R})},\nonumber\\
% 	&\pm t^3\frac{\Delta_\text{L}^2+\Delta_\text{R}^2}{32\Delta_\text{L}^2\Delta_\text{R}^2}\cos\frac{\Delta\Phi}{2}\left(\cos\Delta\Phi-1\right),\\
	E_{2\pm}=&+|\Delta_\text{L}-\Delta_\text{R}|
	\pm\frac{t}{2}\cos\frac{\Delta\Phi}{2}\\
	&-t^2\frac{(\Delta_\text{L}-\Delta_\text{R})^2(1-\cos\Delta\Phi)-\Delta_\text{L}\Delta_\text{R}(1+\cos\Delta\Phi)}{16\Delta_\text{L}\Delta_\text{R}(\Delta_\text{L}+\Delta_\text{R})},\nonumber\\
% 	&\pm t^3\frac{\Delta_\text{L}^2+\Delta_\text{R}^2}{32\Delta_\text{L}^2\Delta_\text{R}^2}\cos\frac{\Delta\Phi}{2}\left(\cos\Delta\Phi-1\right),\\
	E_{3\pm}=&+|\Delta_\text{L}+\Delta_\text{R}|
	\pm\frac{t}{2}\cos\frac{\Delta\Phi}{2}\\
	&-t^2\frac{(\Delta_\text{L}+\Delta_\text{R})^2(1-\cos\Delta\Phi)+\Delta_\text{L}\Delta_\text{R}(1+\cos\Delta\Phi)}{16\Delta_\text{L}\Delta_\text{R}(\Delta_\text{L}+\Delta_\text{R})}.\nonumber%\\
% 	&\pm t^3\frac{\Delta_\text{L}^2+\Delta_\text{R}^2}{32\Delta_\text{L}^2\Delta_\text{R}^2}\cos\frac{\Delta\Phi}{2}\left(\cos\Delta\Phi-1\right),
\end{eqnarray}
For each of these eigenenergies, there are two degenerate eigenstates with an even and odd number of electrons on the quantum dots, respectively. Equivalently, for these states the nonlocal fermion $f=(\gamma^\text{L}_{1A}+i\gamma^\text{R}_{2B})/2$ is occupied or empty.
We remark that the above expressions are only valid for $\Delta_\text{L}\neq\Delta_\text{R}$. For $\Delta_\text{L}=\Delta_\text{R}$, we instead find $E_{1\pm}=0$ and $E_{2\pm}=\pm t\cos(\Delta\Phi/2)$, respectively. Interestingly, the above analytic expressions are in good agreement with the numerically obtained eigenvalues shown in \fref{fig:energies} even for large tunnel couplings $t\sim\Delta$.

In order to see whether a fractional Josephson effect can occur in the system, we have to analyze the phase dependence of the eigenenergies $E_{0\pm}$. As shown in \fref{fig:energies}, the eigenenergies cross at $\Delta\Phi=(2n+1)\pi$, $n\in\mathbb{Z}$. Due to these crossings, the system stays, e.g., in one of the two states with energy $E_{0+}$ when the phase difference is adiabatically increased and passes through $\Delta\Phi=(2n+1)\pi$. Hence, increasing the phase difference by $2\pi$ exchanges the ground state and the first excited state, thus leading to a $4\pi$ periodic Josephson current.
We remark that the observation of the $4\pi$ periodicity requires a conservation of fermion parity over a time scale longer than the time needed to increase the phase difference by $4\pi$.

To understand the origin of these crossings, we take a look at the Majorana representation of the quantum dots, cf. \fref{fig:Majorana}. For any phase difference, there are two uncoupled Majorana modes at the end of the quantum dot chain. They are associated with the degeneracy of the even and odd parity states. In addition, for $\Delta\Phi=(2n+1)\pi$, the system hosts two chains consisting of three Majorana fermions. As these chains are described by an antisymmetric $3\times3$ matrix, they must have a zero-energy mode associated with the level crossing. The zero-energy modes are given by
\begin{eqnarray}
	\Gamma_1=\frac{1}{\sqrt{\Delta^2+t^2/4}}\left(\pm\frac{t}{2}\gamma^\text{L}_{1B}-\Delta\gamma^\text{R}_{1A}\right),\\
	\Gamma_2=\frac{1}{\sqrt{\Delta^2+t^2/4}}\left(\pm\frac{t}{2}\gamma^\text{R}_{2A}-\Delta\gamma^\text{L}_{2B}\right),
\end{eqnarray}
where the upper (lower) sign applies at $\Delta\Phi=(4n\pm1)\pi$. Hence, the zero-energy states are localized at the ends of the three-Majorana chains. The nonlocal fermion $f=(\Gamma_1+i\Gamma_2)/2$ associated with these two modes is occupied (empty) for the states with energy $E_{0\pm}$.

\subsection{\label{ssec:perturbation}Influence of perturbations}
\begin{figure}
	\includegraphics[width=\columnwidth]{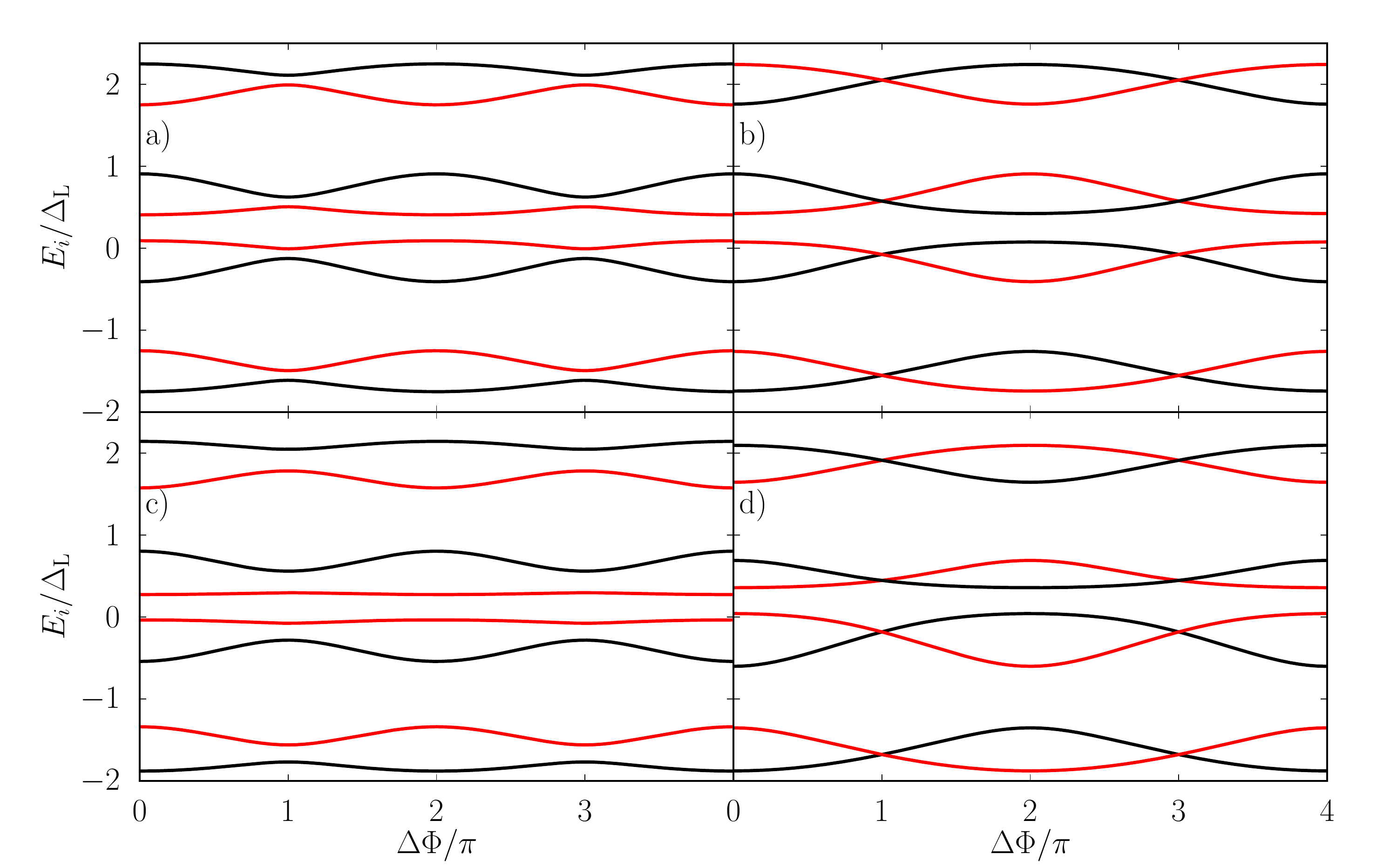}
	\caption{\label{fig:perturbation}Influence of level detunings (a) and b)) and Coulomb interactions (c) and d)) on the eigenenergies of the quadruple quantum dot. In a) we have $\varepsilon_{\text{L}1}=\Delta_\text{L}/2$, in b) $\varepsilon_{\text{L}2}=\Delta_\text{L}/2$, in c) $U_{\text{L}1\text{L}2}=\Delta_\text{L}/2$ and in d) $U_{\text{L}2\text{R}1}=\Delta_\text{L}/2$. Other parameters are as in \fref{fig:energies}.}
\end{figure}
So far, we considered an ideal system tuned to its sweet spot. We now analyze the influence of perturbations away from this point. We discuss both the effect of level detunings as well as of Coulomb interactions.

\subsubsection{Level detunings}
In the following, we analyse the influence of level detunings on the occurrence of the fractional Josephson effect. We first consider the detuning of the outer dot level $\text{L}1$ and $\text{R}2$. As can be seen in \fref{fig:perturbation}a), a detuning of one of the outer dots turns the crossings between $E_{0\pm}$ into anticrossings and, hence, destroys the fractional Josephson effect. The reason for this becomes immediately clear when looking at the Majorana representation of the quantum dots. A finite energy $\varepsilon_{\text{L}1}$ induces a coupling between the zero-energy Majorana modes $\gamma^\text{L}_{1A}$ and $\Gamma_1$. This coupling moves both states to finite energy and, thereby, destroys the level crossing at $\Delta\Phi=(2n+1)\pi$.

We next turn to a detuning of the inner dot levels $\varepsilon_{\text{L}2}$ and $\varepsilon_{\text{R}1}$. In contrast to a detuning of the outer dot levels, we now find that the crossings are preserved and the fractional Josephson effect persists, cf. \fref{fig:perturbation}b). This can again be understood from the Majorana representation of the quantum dots. While a finite $\varepsilon_{\text{L}2}$ induces a coupling between $\gamma^\text{L}_{2A}$ and $\gamma^\text{L}_{2B}$, it does not affect $\Gamma_1$ which therefore stays a zero-energy excitation. More generally, for a detuning of both inner dots, we find the new zero-energy Majorana modes
\begin{eqnarray}
	\Gamma_1=\frac{2}{\sqrt{4\Delta^2+t^2+\varepsilon_{\text{R}1}^2}}\left(\pm\frac{t}{2}\gamma^\text{L}_{1B}-\Delta\gamma^\text{R}_{1A}+\frac{\varepsilon_{\text{R}1}}{2}\gamma^\text{R}_{2A}\right),\\
	\Gamma_2=\frac{2}{\sqrt{4\Delta^2+t^2+\varepsilon_{\text{L}2}^2}}\left(\pm\frac{t}{2}\gamma^\text{R}_{2A}-\Delta\gamma^\text{L}_{2B}+\frac{\varepsilon_{\text{L}2}}{2}\gamma^\text{L}_{1B}\right).
\end{eqnarray}

\subsubsection{Coulomb interactions}
We next turn to the influence of Coulomb interactions on the level crossings. As shown in \fref{fig:perturbation}c), Coulomb interactions within a double quantum dot destroy the level crossings at $\Delta\Phi=(2n+1)\pi$. This is due to the fact that the Coulomb interaction induces couplings between the zero-energy modes of the unperturbed system in analogy to the detuning of the outer dot levels discussed above.

In contrast, Coulomb interactions $U_{\text{L}2\text{R}1}$ between the double dots do not affect the occurrence of the fractional Josephson effect, see \fref{fig:perturbation}d). This can be most easily understood by analyzing the system in the many-body picture. The eigenstates of the two uncoupled double dots are given by the product states
$\ket{\alpha_\pm}_\text{L}\otimes \ket{\alpha_\pm}_\text{R}$, $\ket{\alpha_\pm}_\text{L}\otimes \ket{\beta_\pm}_\text{R}$, $\ket{\beta_\pm}_\text{L}\otimes \ket{\alpha_\pm}_\text{R}$, $\ket{\beta_\pm}_\text{L}\otimes \ket{\beta_\pm}_\text{R}$ where the eigenstates of the double dots are~\cite{leijnse_parity_2012}
\begin{eqnarray}
	\ket{\alpha_\pm}_r=\frac{1}{\sqrt{2}}\left(e^{-i\Phi_r/2}\ket{00}_r\pm e^{i\Phi_r/2}\ket{11}_r\right),\quad E_{\alpha_\pm}=\pm \Delta_r,\\
	\ket{\beta_\pm}_r=\frac{1}{\sqrt{2}}\left(\ket{01}_r\pm\ket{10}_r\right),\quad E_{\beta_\pm}=\pm t_r,
\end{eqnarray}
with $\ket{11}\equiv c_{r1}^\dagger c_{r2}^\dagger \ket{00}$. The fourfold degenerate ground states at the sweet spot are given by $\ket{\phi_0}=\ket{\alpha_-}_\text{L}\otimes\ket{\alpha_-}_\text{R}$, $\ket{\phi_1}=\ket{\alpha_-}_\text{L}\otimes\ket{\beta_-}_\text{R}$, $\ket{\phi_2}=\ket{\beta_-}_\text{L}\otimes\ket{\alpha_-}_\text{R}$ and $\ket{\phi_3}=\ket{\beta_-}_\text{L}\otimes\ket{\beta_-}_\text{R}$. 

To first order in perturbation theory, the Coulomb interaction between the double dots does not mix these states. In addition, it yields the same energy shift for all four states. Hence, the level crossing at $\Delta\Phi=(2n+1)\pi$ is conserved. As can be seen from the numerical analysis in \fref{fig:perturbation}d), the argument remains valid even for large Coulomb interactions where lowest-order perturbation theory is no longer reliable.

\section{\label{sec:strong}Strong-coupling limit}
\begin{figure}
	\includegraphics[width=.5\columnwidth]{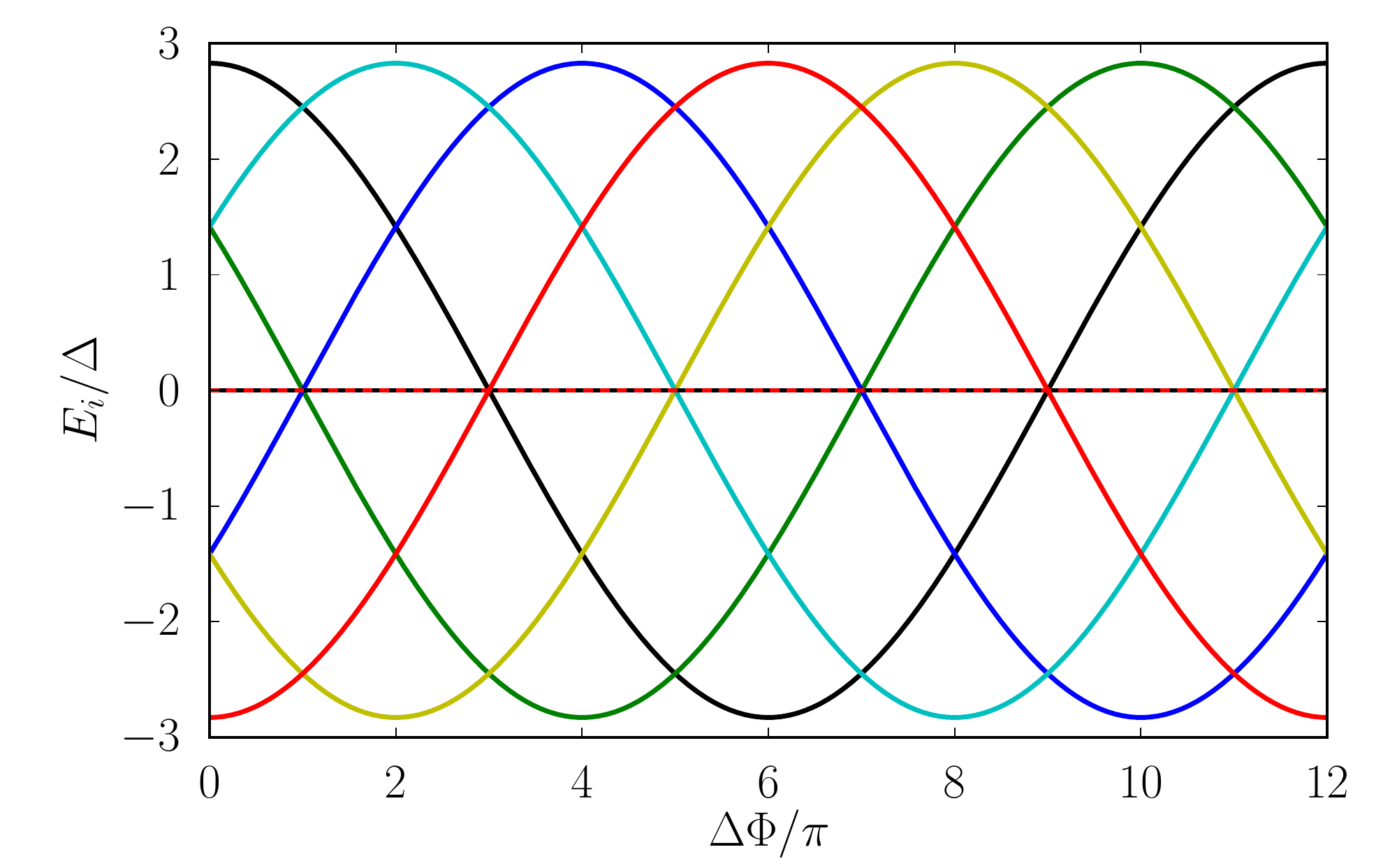}
	\includegraphics[width=.5\columnwidth]{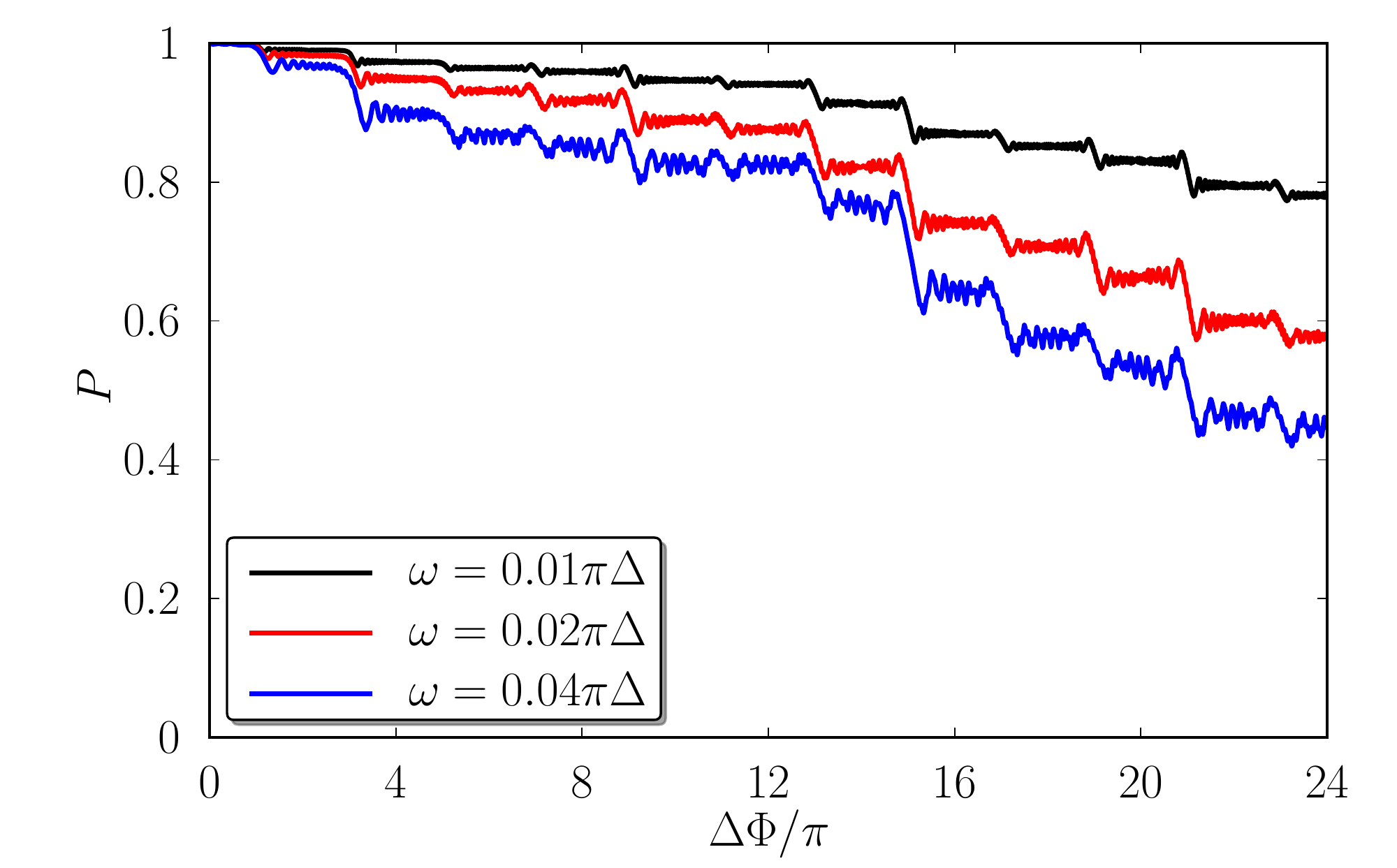}
	\caption{\label{fig:Especial1}Left: Eigenenergies as a function of phase difference for a symmetric setup $\Delta_\text{L}=\Delta_\text{R}\equiv\Delta$ with $\varepsilon_{ri}=U_{rir'i'}=0$ and $t=\sqrt{2}\Delta$. Right: Probability $P$ to find the system in the state with energy $E=-2\sqrt{2}\Delta\cos\Delta(\Phi/6)$ for a time-dependent phase difference $\Delta\Phi(\tau)=\omega\tau$.}
\end{figure}
\begin{figure}
	\includegraphics[width=.5\columnwidth]{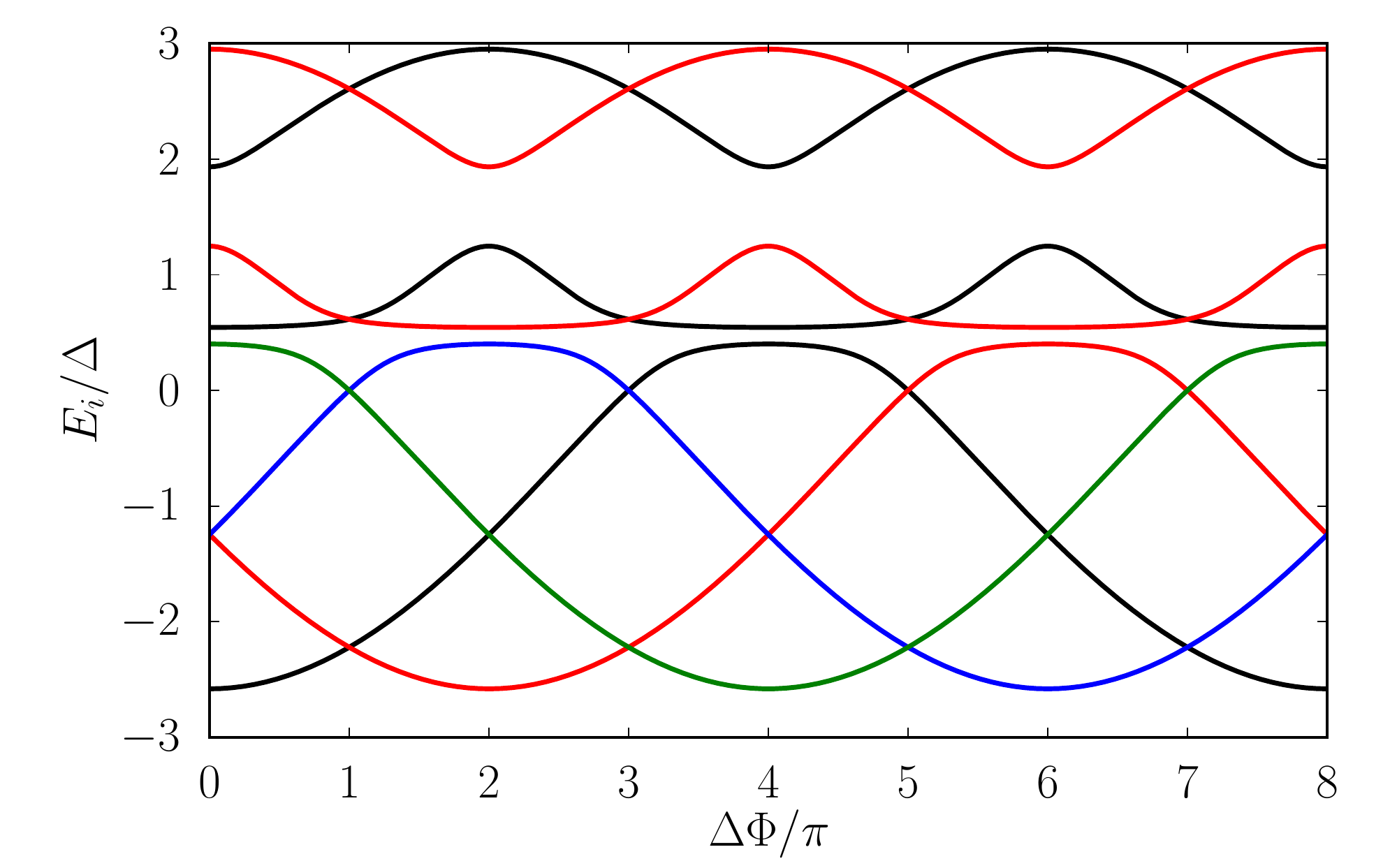}
	\includegraphics[width=.5\columnwidth]{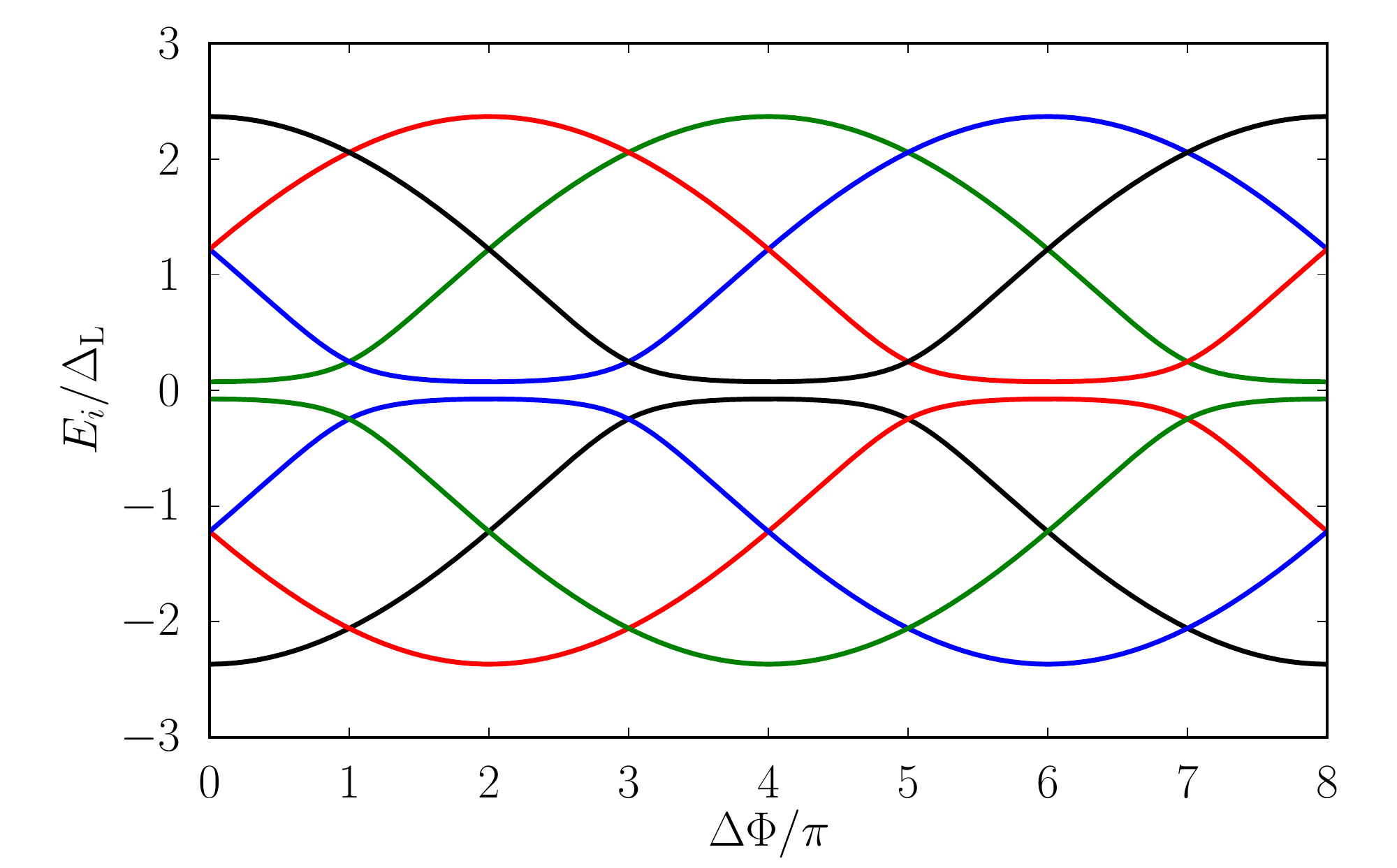}
	\caption{\label{fig:Especial2}Eigenenergies as a function of phase difference. Left: Symmetric system with Coulomb interaction between the double dots $U_{\text{L}2\text{R}1}=\Delta$ and $t=1.24697\Delta$. Other parameters as in \fref{fig:Especial1}. Right: Asymmetric system $\Delta_\text{R}=0.7\Delta_\text{L}$ with $t=1.147\Delta_\text{L}$. Other parameters as in \fref{fig:Especial1}.}
\end{figure}

As the tunnel coupling between the double dots is increased, the splitting between the energies $E_{i\pm}$ at $\Delta\Phi=2n\pi$ increases as well. As was first pointed out in Ref.~\cite{nogueira_strong-coupling_2012}, for a symmetric system $\Delta_\text{L}=\Delta_\text{R}\equiv\Delta$ tuned to the sweet spot, this increased splitting leads to the occurrence of new level crossings in the spectrum at $t=\sqrt{2}\Delta$. While this strong-coupling regime might be hard to realize in a nanowire setup, it should be easily realizable in a quantum-dot system. In the following, we demonstrate that these additional crossings give rise to a $12\pi$-periodic Josephson effect. We then discuss the robustness of this effect with respect to asymmetries in the systems and perturbations away from the sweet spot.

The spectrum of a symmetric system $\Delta_\text{L}=\Delta_\text{R}$ with $t=\sqrt{2}\Delta$ is shown in \fref{fig:Especial1}. In addition to the level crossings at $\Delta\Phi=(2n+1)\pi$, new crossings at $\Delta\Phi=2n\pi$ arise, leading to eigenenergies that are $12\pi$ periodic in the phase difference. We emphasize that these additional crossings are of accidential nature and not associated with Majorana degrees of freedom. In order to demonstrate that the system can exhibit a $12\pi$ Josephson effect in the strong-coupling limit, we consider the following situation. Let us assume that at time $\tau=0$ the system has a phase difference $\Delta\Phi=0$ and is prepared in its ground state. We then increase the phase difference with time according to $\Delta\Phi(\tau)=\omega\tau$. By numerically solving the time-dependent Schrödinger equation, we calculate the probability $P$ to find the system in the state with energy $E=-2\sqrt{2}\Delta\cos(\Delta\Phi/6)$. The results shown in \fref{fig:Especial1} clearly indicate that for slow driving, the system indeed stays on one energy branch and, thus, exhibits a $12\pi$-periodic Josephson effect.

We finally address the robustness of the additional crossings. As they are of accidential nature, any perturbation will destroy them. This prevents the occurrence of the $12\pi$ Josephson effect and leads to the $4\pi$ effect discussed above. Importantly, there is however the possibility to get at least some of the additional crossings in the presence of perturbations.
First of all, additional crossings arise in a symmetric system with Coulomb interaction $U_{\text{L}2\text{R}1}$ between the two double dots. These give rise to $8\pi$-periodic eigenenergies, cf. \fref{fig:Especial2}. Furthermore, additional crossings can occur for an asymmetric system $\Delta_\text{L}\neq\Delta_\text{R}$ tuned to the sweet spot. As shown in \fref{fig:Especial2}, here again, the additional crossings lead to the occurrence of an $8\pi$ Josephson effect.

\section{\label{sec:conclusions}Conclusions}
We investigated the Josephson current through a system of two tunnel-coupled double quantum dots subject to an inhomogenous magnetic field. Similarly to a Majorana nanowire, the system exhibits a crossing of energy levels that gives rise to a $4\pi$ Josephson effect. We find this effect to be robust with respect to a detuning of the inner dot levels as well as to Coulomb interactions between the double dots. In contrast, detunings of the outer dot levels and Coulomb interactions within a double dot convert the crossings into anticrossing and lead to a conventional $2\pi$-periodic effect.
In addition, we investigated the regime of strongly couplgy double dots. We found that in this situation additional accidential level crossings can occur. For a symmetric setup, they lead to a $12\pi$-periodic Josephson effect whereas for an asymmetric system an $8\pi$-periodic effect arises. Both effects require a fine-tuning of parameters and are vulnerable to perturbations.

\ack
We thank Martin Leijnse for reading the manuscript and useful feedback. We acknowledge financial support from the Swiss NSF and NCCR QSIT.

\appendix
\section{\label{app:Heff}Derivation of the effective dot Hamiltonian}
In the following, we derive the effective dot Hamiltonian~\eref{eq:Heff}. For simplicity, we suppress the index $r=\mathsf{L,R}$ characterizing the two double dots. In the limit of an infinite superconducting gap, the superconductor's degree of freedom can be integrated out exactly to yield an effective dot Hamiltonian~\cite{eldridge_superconducting_2010}
\begin{equation}
	H=\sum_i H_i+H_\text{int}+H_\text{tun}+H_\text{prox}.
\end{equation}
Here, 
\begin{equation}\label{eq:Hi}
	H_i=\sum_{i\sigma} \varepsilon_{i} n_{i\sigma}+\sum_i\vec B_i\cdot\vec S_i
\end{equation}
describes the relevant single level with energy $\varepsilon_i$ in each quantum dot as well as the influence of an externally applied magnetic field $\vec B_i$ acting on the dot spin $\vec S_i=\sum_{\sigma\sigma'}\frac{1}{2}c_{i\sigma}^\dagger \boldsymbol \sigma_{\sigma\sigma'} c_{i\sigma'}$.

The effects of interdot Coulomb interactions $U_i$ as well as of intradot Coulomb interactions $U$ are described by
\begin{equation}\label{eq:Hint}
	H_\text{int}=\sum_i U_in_{i\up}n_{i\down}+\sum_{\sigma\sigma'} U n_{1\sigma}n_{2\sigma'}.
\end{equation}
Tunneling between the two dots is characterized by the tunneling Hamiltonian
\begin{equation}\label{eq:Htun}
	H_\text{tun}=\sum_\sigma \tilde t c_{1\sigma}^\dagger c_{2\sigma}+\text{H.c.}
\end{equation}
Here, we assumed that tunneling conserves the electron spin, i.e., there is no spin-orbit interaction in the quantum dots.

Finally, the superconducting proximity effect on the double dot is captured by
\begin{equation}\label{eq:Hprox}
	H_\text{prox}=-\sum_i \frac{\Gamma_{Si}}{2}\left(c_{i\up}^\dagger c_{i\down}^\dagger+\text{H.c.}\right)+\frac{\Gamma_\text{S}}{2}\left(c_{2\up}^\dagger c_{1\down}^\dagger-c_{2\down}^\dagger c_{1\up}^\dagger+\text{H.c.}\right),
\end{equation}
where $\Gamma_\text{S}=\sqrt{\Gamma_\text{S1}\Gamma_\text{S2}}$ is the geometric average of the tunnel couplings between the superconductor and dot 1 and 2, respectively. The first term describes a local proximity effect on a single dot whereas the second term captures the nonlocal proximity effect that involves both dots.

In the following, we focus on the situation where a large magnetic field is applied to the double dot such that for each dot only a single spin state is relevant for transport. This directly implies that double occupancy of each dot is forbidden and a local proximity effect cannot occur. In the case of a homogenous magnetic field, we conclude from~\eref{eq:Hprox} that no proximity effect is possible at all since it requires the pairing of electrons with antiparallel spin. If the magnetic field is pointing in opposite directions in the two dots, we similarly conclude that no tunneling between the two double dots is possible as it conserves spin. For an arbitrary angle $\phi$ enclosed between $\vec B_1$ and $\vec B_2$, the effective tunnel coupling is given by $t=\tilde t\cos\phi/2$ while the effective strength of the proximity effect is characterized by $\Delta=\frac{\Gamma_\mathsf{S}}{2}\sin\phi/2$. As pointed out in~\cite{leijnse_parity_2012}, this tunability of the ratio between $t$ and $\Delta$ is crucial to achieve the special point $t=\Delta$ required for the occurrence of Majorana fermions on the double dot. We thus conclude that each double quantum dot $r$ can be described by an effective dot Hamiltonian of the form
\begin{equation}
	H=\sum_i\varepsilon_i n_i+U n_1 n_2+t c_1^\dagger c_2+\Delta e^{i\Phi} c_1^\dagger c_2^\dagger+\text{H.c.},
\end{equation}
where we also included the dependence of the pairing term on the phase $\Phi$ of the superconductor. As a final remark, we note that instead of applying an inhomogenous magnetic field we could also make use of quantum dots with spin-orbit interactions in combination with an externally applied homogenous magnetic field pointing in a different direction than the spin-orbit field.

\section*{References}

% \bibliographystyle{unsrt.bst}
% \bibliographystyle{iopart-num}
% \bibliography{/home/bjoern/LaTeX/Bibtex/Meine_Bibliothek}

\begin{thebibliography}{10}
\expandafter\ifx\csname url\endcsname\relax
  \def\url#1{{\tt #1}}\fi
\expandafter\ifx\csname urlprefix\endcsname\relax\def\urlprefix{URL }\fi
\providecommand{\eprint}[2][]{\url{#2}}

\bibitem{alicea_new_2012}
Alicea J 2012 {\em Rep. Prog. Phys.\/} {\bf 75} 076501 ISSN 0034-4885,
  1361-6633 \urlprefix\url{http://iopscience.iop.org/0034-4885/75/7/076501}

\bibitem{beenakker_search_2013}
Beenakker C 2013 {\em Annu. Rev. Cond. Mat. Phys.\/} {\bf 4} 113--136
  \urlprefix\url{http://www.annualreviews.org/doi/abs/10.1146/annurev-conmatphys-030212-184337}

\bibitem{alicea_non-abelian_2011}
Alicea J, Oreg Y, Refael G, Oppen F~v and Fisher M~P~A 2011 {\em Nature
  Physics\/} {\bf 7} 412--417 ISSN 1745-2473
  \urlprefix\url{http://www.nature.com/nphys/journal/v7/n5/full/nphys1915.html}

\bibitem{flensberg_non-abelian_2011}
Flensberg K 2011 {\em Phys. Rev. Lett.\/} {\bf 106} 090503
  \urlprefix\url{http://link.aps.org/doi/10.1103/PhysRevLett.106.090503}

\bibitem{van_heck_coulomb-assisted_2012}
van Heck B, Akhmerov A~R, Hassler F, Burrello M and Beenakker C~W~J 2012 {\em
  New J. Phys.\/} {\bf 14} 035019 ISSN 1367-2630
  \urlprefix\url{http://iopscience.iop.org/1367-2630/14/3/035019}

\bibitem{moore_nonabelions_1991}
Moore G and Read N 1991 {\em Nucl. Phys. B\/} {\bf 360} 362--396 ISSN 0550-3213
  \urlprefix\url{http://www.sciencedirect.com/science/article/pii/055032139190407O}

\bibitem{kitaev_unpaired_2001}
Kitaev A~Y 2001 {\em Phys.-Usp.\/} {\bf 44} 131--136 ISSN 1468-4780
  \urlprefix\url{http://iopscience.iop.org/1063-7869/44/10S/S29}

\bibitem{das_sarma_proposal_2006}
Das~Sarma S, Nayak C and Tewari S 2006 {\em Phys. Rev. B\/} {\bf 73} 220502
  \urlprefix\url{http://link.aps.org/doi/10.1103/PhysRevB.73.220502}

\bibitem{fu_superconducting_2008}
Fu L and Kane C~L 2008 {\em Phys. Rev. Lett.\/} {\bf 100} 096407
  \urlprefix\url{http://link.aps.org/doi/10.1103/PhysRevLett.100.096407}

\bibitem{fu_josephson_2009}
Fu L and Kane C~L 2009 {\em Phys. Rev. B\/} {\bf 79} 161408
  \urlprefix\url{http://link.aps.org/doi/10.1103/PhysRevB.79.161408}

\bibitem{lutchyn_majorana_2010}
Lutchyn R~M, Sau J~D and Das~Sarma S 2010 {\em Phys. Rev. Lett.\/} {\bf 105}
  077001 \urlprefix\url{http://link.aps.org/doi/10.1103/PhysRevLett.105.077001}

\bibitem{oreg_helical_2010}
Oreg Y, Refael G and von Oppen F 2010 {\em Phys. Rev. Lett.\/} {\bf 105} 177002
  \urlprefix\url{http://link.aps.org/doi/10.1103/PhysRevLett.105.177002}

\bibitem{law_majorana_2009}
Law K~T, Lee P~A and Ng T~K 2009 {\em Phys. Rev. Lett.\/} {\bf 103} 237001
  \urlprefix\url{http://link.aps.org/doi/10.1103/PhysRevLett.103.237001}

\bibitem{flensberg_tunneling_2010}
Flensberg K 2010 {\em Phys. Rev. B\/} {\bf 82} 180516
  \urlprefix\url{http://link.aps.org/doi/10.1103/PhysRevB.82.180516}

\bibitem{badiane_nonequilibrium_2011}
Badiane D~M, Houzet M and Meyer J~S 2011 {\em Phys. Rev. Lett.\/} {\bf 107}
  177002 \urlprefix\url{http://link.aps.org/doi/10.1103/PhysRevLett.107.177002}

\bibitem{beenakker_fermion-parity_2013}
Beenakker C~W~J, Pikulin D~I, Hyart T, Schomerus H and Dahlhaus J~P 2013 {\em
  Phys. Rev. Lett.\/} {\bf 110} 017003
  \urlprefix\url{http://link.aps.org/doi/10.1103/PhysRevLett.110.017003}

\bibitem{ioselevich_anomalous_2011}
Ioselevich P~A and Feigel’man M~V 2011 {\em Phys. Rev. Lett.\/} {\bf 106}
  077003 \urlprefix\url{http://link.aps.org/doi/10.1103/PhysRevLett.106.077003}

\bibitem{jiang_unconventional_2011}
Jiang L, Pekker D, Alicea J, Refael G, Oreg Y and von Oppen F 2011 {\em Phys.
  Rev. Lett.\/} {\bf 107} 236401
  \urlprefix\url{http://link.aps.org/doi/10.1103/PhysRevLett.107.236401}

\bibitem{law_robustness_2011}
Law K~T and Lee P~A 2011 {\em Phys. Rev. B\/} {\bf 84} 081304
  \urlprefix\url{http://link.aps.org/doi/10.1103/PhysRevB.84.081304}

\bibitem{san-jose_ac_2012}
San-Jose P, Prada E and Aguado R 2012 {\em Phys. Rev. Lett.\/} {\bf 108} 257001
  \urlprefix\url{http://link.aps.org/doi/10.1103/PhysRevLett.108.257001}

\bibitem{van_heck_coulomb_2011}
van Heck B, Hassler F, Akhmerov A~R and Beenakker C~W~J 2011 {\em Phys. Rev.
  B\/} {\bf 84} 180502
  \urlprefix\url{http://link.aps.org/doi/10.1103/PhysRevB.84.180502}

\bibitem{dominguez_dynamical_2012}
Domínguez F, Hassler F and Platero G 2012 {\em Phys. Rev. B\/} {\bf 86} 140503
  \urlprefix\url{http://link.aps.org/doi/10.1103/PhysRevB.86.140503}

\bibitem{pikulin_phenomenology_2012}
Pikulin D~I and Nazarov Y~V 2012 {\em Phys. Rev. B\/} {\bf 86} 140504
  \urlprefix\url{http://link.aps.org/doi/10.1103/PhysRevB.86.140504}

\bibitem{sau_possibility_2012}
Sau J~D, Berg E and Halperin B~I 2012 {\em {arXiv:1206.4596}\/}
  \urlprefix\url{http://arxiv.org/abs/1206.4596}

\bibitem{mourik_signatures_2012}
Mourik V, Zuo K, Frolov S~M, Plissard S~R, Bakkers E~P~A~M and Kouwenhoven L~P
  2012 {\em Science\/} {\bf 336} 1003--1007 ISSN 0036-8075, 1095-9203
  \urlprefix\url{http://www.sciencemag.org/content/336/6084/1003}

\bibitem{das_zero-bias_2012}
Das A, Ronen Y, Most Y, Oreg Y, Heiblum M and Shtrikman H 2012 {\em Nature
  Physics\/} {\bf 8} 887--895 ISSN 1745-2473
  \urlprefix\url{http://www.nature.com/nphys/journal/v8/n12/full/nphys2479.html}

\bibitem{deng_anomalous_2012}
Deng M~T, Yu C~L, Huang G~Y, Larsson M, Caroff P and Xu H~Q 2012 {\em Nano
  Lett.\/} {\bf 12} 6414--6419 ISSN 1530-6984
  \urlprefix\url{http://dx.doi.org/10.1021/nl303758w}

\bibitem{finck_anomalous_2013}
Finck A~D~K, Van~Harlingen D~J, Mohseni P~K, Jung K and Li X 2013 {\em Phys.
  Rev. Lett.\/} {\bf 110} 126406
  \urlprefix\url{http://link.aps.org/doi/10.1103/PhysRevLett.110.126406}

\bibitem{churchill_superconductor-nanowire_2013}
Churchill H~O~H, Fatemi V, Grove-Rasmussen K, Deng M~T, Caroff P, Xu H~Q and
  Marcus C~M 2013 {\em Phys. Rev. B\/} {\bf 87} 241401
  \urlprefix\url{http://link.aps.org/doi/10.1103/PhysRevB.87.241401}

\bibitem{bagrets_class_2012}
Bagrets D and Altland A 2012 {\em Phys. Rev. Lett.\/} {\bf 109} 227005
  \urlprefix\url{http://link.aps.org/doi/10.1103/PhysRevLett.109.227005}

\bibitem{pikulin_zero-voltage_2012}
Pikulin D~I, Dahlhaus J~P, Wimmer M, Schomerus H and Beenakker C~W~J 2012 {\em
  New J. Phys.\/} {\bf 14} 125011 ISSN 1367-2630
  \urlprefix\url{http://iopscience.iop.org/1367-2630/14/12/125011}

\bibitem{rokhinson_fractional_2012}
Rokhinson L~P, Liu X and Furdyna J~K 2012 {\em Nature Physics\/} {\bf 8}
  795--799 ISSN 1745-2473
  \urlprefix\url{http://www.nature.com/nphys/journal/v8/n11/full/nphys2429.html}

\bibitem{williams_unconventional_2012}
Williams J~R, Bestwick A~J, Gallagher P, Hong S~S, Cui Y, Bleich A~S, Analytis
  J~G, Fisher I~R and Goldhaber-Gordon D 2012 {\em Phys. Rev. Lett.\/} {\bf
  109} 056803
  \urlprefix\url{http://link.aps.org/doi/10.1103/PhysRevLett.109.056803}

\bibitem{sau_realizing_2012}
Sau J~D and Sarma S~D 2012 {\em Nat. Commun.\/} {\bf 3} 964 ISSN 2041-1723
  \urlprefix\url{http://www.nature.com/ncomms/journal/v3/n7/full/ncomms1966.html}

\bibitem{leijnse_parity_2012}
Leijnse M and Flensberg K 2012 {\em Phys. Rev. B\/} {\bf 86} 134528
  \urlprefix\url{http://link.aps.org/doi/10.1103/PhysRevB.86.134528}

\bibitem{fulga_adaptive_2013}
Fulga I~C, Haim A, Akhmerov A~R and Oreg Y 2013 {\em New J. Phys.\/} {\bf 15}
  045020 ISSN 1367-2630
  \urlprefix\url{http://iopscience.iop.org/1367-2630/15/4/045020}

\bibitem{wright_localized_2013}
Wright A~R and Veldhorst M 2013 {\em {arXiv:1303.1570}\/}
  \urlprefix\url{http://arxiv.org/abs/1303.1570}

\bibitem{brunetti_anomalous_2013}
Brunetti A, Zazunov A, Kundu A and Egger R 2013 {\em {arXiv:1305.3816}\/}
  \urlprefix\url{http://arxiv.org/abs/1305.3816}

\bibitem{hofstetter_cooper_2009}
Hofstetter L, Csonka S, Nygard J and Schönenberger C 2009 {\em Nature\/} {\bf
  461} 960--963 ISSN 0028-0836
  \urlprefix\url{http://dx.doi.org/10.1038/nature08432}

\bibitem{herrmann_carbon_2010}
Herrmann L~G, Portier F, Roche P, Levy~Yeyati A, Kontos T and Strunk C 2010
  {\em Phys. Rev. Lett.\/} {\bf 104} 026801 copyright (C) 2010 The American
  Physical Society; Please report any problems to prola@aps.org
  \urlprefix\url{http://link.aps.org/abstract/PRL/v104/e026801}

\bibitem{hofstetter_finite-bias_2011}
Hofstetter L, Csonka S, Baumgartner A, Fülöp G, {d’Hollosy} S, Nygård J
  and Schönenberger C 2011 {\em Phys. Rev. Lett.\/} {\bf 107} 136801
  \urlprefix\url{http://link.aps.org/doi/10.1103/PhysRevLett.107.136801}

\bibitem{das_high-efficiency_2012}
Das A, Ronen Y, Heiblum M, Mahalu D, Kretinin A~V and Shtrikman H 2012 {\em
  Nat. Commun.\/} {\bf 3} 1165 ISSN 2041-1723
  \urlprefix\url{http://www.nature.com/ncomms/journal/v3/n10/full/ncomms2169.html}

\bibitem{schindele_near-unity_2012}
Schindele J, Baumgartner A and Schönenberger C 2012 {\em Phys. Rev. Lett.\/}
  {\bf 109} 157002
  \urlprefix\url{http://link.aps.org/doi/10.1103/PhysRevLett.109.157002}

\bibitem{van_dam_supercurrent_2006}
van Dam J~A, Nazarov Y~V, Bakkers E~P~A~M, De~Franceschi S and Kouwenhoven L~P
  2006 {\em Nature\/} {\bf 442} 667--670 ISSN 0028-0836
  \urlprefix\url{http://dx.doi.org/10.1038/nature05018}

\bibitem{cleuziou_carbon_2006}
Cleuziou J~P, Wernsdorfer W, Bouchiat V, Ondarcuhu T and Monthioux M 2006 {\em
  Nat. Nano.\/} {\bf 1} 53--59 ISSN 1748-3387
  \urlprefix\url{http://dx.doi.org/10.1038/nnano.2006.54}

\bibitem{grove-rasmussen_kondo_2007}
Grove-Rasmussen K, Jørgensen H~I and Lindelof P~E 2007 {\em New J. Phys.\/}
  {\bf 9} 124--124 ISSN 1367-2630
  \urlprefix\url{http://iopscience.iop.org/1367-2630/9/5/124/}

\bibitem{winkelmann_superconductivity_2009}
Winkelmann C~B, Roch N, Wernsdorfer W, Bouchiat V and Balestro F 2009 {\em Nat.
  Phys.\/} {\bf 5} 876--879 ISSN 1745-2473
  \urlprefix\url{http://dx.doi.org/10.1038/nphys1433}

\bibitem{eichler_tuning_2009}
Eichler A, Deblock R, Weiss M, Karrasch C, Meden V, Schönenberger C and
  Bouchiat H 2009 {\em Phys. Rev. B\/} {\bf 79} 161407 copyright (C) 2010 The
  American Physical Society; Please report any problems to prola@aps.org
  \urlprefix\url{http://link.aps.org/abstract/PRB/v79/e161407}

\bibitem{kanai_electrical_2010}
Kanai Y, Deacon R~S, Oiwa A, Yoshida K, Shibata K, Hirakawa K and Tarucha S
  2010 {\em Phys. Rev. B\/} {\bf 82} 054512
  \urlprefix\url{http://link.aps.org/doi/10.1103/PhysRevB.82.054512}

\bibitem{lee_zero-bias_2012}
Lee E~J~H, Jiang X, Aguado R, Katsaros G, Lieber C~M and De~Franceschi S 2012
  {\em Phys. Rev. Lett.\/} {\bf 109} 186802
  \urlprefix\url{http://link.aps.org/doi/10.1103/PhysRevLett.109.186802}

\bibitem{kim_transport_2013}
Kim B~K, Ahn Y~H, Kim J~J, Choi M~S, Bae M~H, Kang K, Lim J~S, López R and Kim
  N 2013 {\em Phys. Rev. Lett.\/} {\bf 110} 076803
  \urlprefix\url{http://link.aps.org/doi/10.1103/PhysRevLett.110.076803}

\bibitem{beenakker_three_1992}
Beenakker C~W~J 1992 Three "universal" mesoscopic josephson effects {\em
  Transport Phenomena in Mesoscopic Systems\/} ed Fukuyama H and Ando T
  (Berlin: Springer)

\bibitem{nogueira_strong-coupling_2012}
Nogueira F~S and Eremin I 2012 {\em J. Phys.: Condens. Matter\/} {\bf 24}
  325701 ISSN 0953-8984, 1361-{648X}
  \urlprefix\url{http://iopscience.iop.org/0953-8984/24/32/325701}

\bibitem{eldridge_superconducting_2010}
Eldridge J, Pala M~G, Governale M and König J 2010 {\em Phys. Rev. B\/} {\bf
  82} 184507 \urlprefix\url{http://link.aps.org/doi/10.1103/PhysRevB.82.184507}

\end{thebibliography}

\providecommand{\newblock}{}

\end{document}